\documentclass[doublecol]{epl2}
\usepackage{amsmath, amssymb, hyperref}

\title{Lorentz-preserving fields in Lorentz-violating theories}

\author{O. Ganguly\inst{1}\thanks{oindrila@bose.res.in} \and D. Gangopadhyay\inst{1}\thanks{debashis@bose.res.in} \and P. Majumdar\inst{2}\thanks{parthasarathi.majumdar@saha.ac.in}}
\shortauthor{O. Ganguly \etal}
\institute{
	\inst{1}S. N. Bose National Centre for Basic Sciences, Kolkata 700098, India\\
	\inst{2}Saha Institute of Nuclear Physics, Kolkata 700064, India
}

\abstract{
We identify a fairly general class of field configurations (of spins
$0, \frac{1}{2}$ and $1$) which preserve Lorentz invariance in effective field
theories of Lorentz violation characterized by a constant timelike
vector. These fields concomitantly satisfy the equations of motion yielding
cubic dispersion relations similar to those found earlier. They appear to have
prospective applications in inflationary scenarios.
}

\begin{document}

\maketitle

\section{Introduction}

Invariance under Lorentz transformation is known till date to be a global
symmetry of the standard theory of elementary particles when gravitation is
ignored. However, questions have been raised regarding the
validity of this symmetry at small length scales owing to
probable quantum gravity effects \cite{carroll90, latorre95, colladay97, amelino98, colladay98, coleman99, gambini99, schaefer99, biller99, kifune99, amelino99, ellis00, urrutia00, piran01, kostelecky01, kostelecky02, sarkar02, alfaro02, alfaro026, ellis03, jacob03, jacob05, albert07, albert072, albert08, piran07, galaverni07, gibbons07, liberati08, maccione08, kostelecky08, kostelecky09, ellis09, smolin09, abdo09, amelino09, gubitosi10}. The natural mass scale of quantum gravity is the Planck mass $M_{Pl}$. Departures, suppressed by the Planck mass, from the
standard special relativistic dispersion relation of free particles of mass
$m$ at large energies have been accepted as a signature of Lorentz invariance
violation and has been the principal \textit{objet de l'attention} of experimental
and theoretical probes of Lorentz violation. These hypothesised \textit{ad hoc} corrections
due to Lorentz non-invariance must have their origin in new terms in the
action of the system. Myers and Pospelov \cite{mp03} have studied this issue
within the framework of effective field theory involving fields of spins 0,
1/2 and 1, by incorporating into the action dimension five operators containing a
{\it constant} timelike 4-vector $\mathbf{n}$ which ostensibly breaks Lorentz
invariance. Choosing a Lorentz frame where $n^\mu = (1, \vec{0})$,
corrections of $O(p^3)$ to the dispersion relation of each of the three fields
have been obtained in \cite{mp03} in the limit of relatively high energies $E$ 
($M_{Pl}>> E >> m$). 

For a complex scalar field this is given by
\begin{align}
\omega^2 \simeq |\vec{p}|^2 + m^2 + \frac{\kappa}{M_{Pl}}|\vec{p}|^3. \label{eq:scalardisp}
\end{align} 
For the Maxwell field, the dispersion relation obtained takes the form (for
circularly polarized photons) \cite{mp03, jacob05}
\begin{align}
\omega_{R,L}^2 \simeq |\vec{p}|^2 \pm \frac{2\xi}{M_{Pl}}|\vec{p}|^3 
\label{eq:vecdisp}
\end{align}


In the case of a Dirac spinor one gets \cite{mp03},
\begin{align}
\left[\omega^2 - |\vec{p}|^2 - m^2 - \frac{2 |\vec{p}|^3}{M_{PL}}(\eta_1 + \eta_2 \gamma_5)\right] 
\psi &\simeq 0 , \nonumber \\
\omega^2 - |\vec{p}|^2 - m^2 - \frac{2 |\vec{p}|^3}{M_{PL}}\eta_{R,L} 
&\simeq 0. 
\label{eq:spindisp}
\end{align}
In (\ref{eq:spindisp}), the spinors have been chosen to be eigenstates of the chirality operator which is a valid assumption at high energies. $\eta_{R,L} \equiv \eta_1 \pm \eta_2$. 

Many experiments aimed at constraining the parameters $\kappa, \xi, \eta_1, \eta_2$
quantifying Lorentz violation have been proposed in the past few years. Lorentz
violating effects scale with energy making astrophysical observations a
perfect arena for detecting them. The simplest astrophysical observations that
provide interesting constraints on lack of Lorentz symmetry at Planck scale
measure the differences in arrival times of photons emitted simultaneously
from distant sources of radiation like $\gamma$-ray bursts, active galactic
nuclei and pulsars \cite{amelino98, schaefer99, biller99, ellis00, ellis03, galaverni07, liberati08, ellis09}. The authors of \cite{albert07, albert072, albert08} found stringent limits on $\xi$ by recording the timing of photons emitted during strong flares of the active galactic nucleus
Markarian 501. The lowest order corrections in the photon dispersion relation
(\ref{eq:vecdisp}) also imply the birefringence of vacuum (different group
velocities for different helicities of photons). In 2008, Maccione \textit{et
al.} \cite{maccione08} used polarimetric observations of hard x-ray from the
Crab nebula to impose a bound on Lorentz violation in quantum electrodynamics
of $|\xi| < 9 \times 10^{-10}$ at $95\% $ confidence level.

Complementary constraints have also been obtained from the threshold reactions
of photon decay, fermion pair emission, synchrotron radiation, vacuum Cerenkov
radiation and helicity changing decays. In \cite{jacob03}, the authors analyzed
synchrotron radiation from the Crab nebula to deduce $\eta > -7 \times
10^{-8}$. Observational details and their phenomenological consequences have
been exhaustively discussed in \cite{matt05, jacob06, liberati09}.   

It is clear that the deformed dispersion relation has been the object of
extensive observational scrutiny of departure from Lorentz invariance. Does it
unequivocally imply Lorentz violation? We first explore the possibility that
special field configurations exist for which the apparently Lorentz symmetry
violating action of \cite{mp03} may still be Lorentz \textit{invariant} in close analogy
to what happens in magnetic monopole theory, as shown by Zwanziger in
\cite{zwan71}. In Zwanziger's work, a local, manifestly anisotropic Lagrangian
density has been shown to preserve Lorentz invariance when the fields
obey certain constraints. In this paper we consider the N\"{o}ther current corresponding
to Lorentz transformation for the higher derivative theory proposed in
\cite{mp03}. Requiring that this N\"{o}ther current is conserved leads to the the condition that when the fields are decomposed in a particular way, the effective action remains Lorentz invariant. The initial absence of Lorentz symmetry in the action is transferred to the Lorentz non-invariant splitting of the fields. Identical configurations also appear
when we demand that the \textit{action} changes at most by a constant when the
fields transform under an infinitesimal Lorentz transformation while the
$4$-vector $\mathbf{n}$ stays fixed. We further investigate the dispersion relation that these special field configurations satisfy.

This particular decomposition may have direct application in the physics of the early Universe when energies were sufficiently high for Lorentz violation to have been present. Usually inhomogeinities are related to density perturbations
which act as seeds for structure formation. In this commonly practised approach , there is no
specific mention of quantum gravity effects that lead to inhomogeneities. In fact, the metric fluctuations responsible for the growth of large scale anisotropies originate from quantum fluctuations of the inflaton field. We give an illustration of how Lorentz invariance 
violation near the Big Bang can be related to inhomogeneous Lorentz preserving fields which in 
turn might give rise to structure formation. This scenario will be developed in 
the context of the  Myers and Pospelov \cite{mp03} model. 

\section{Lorentz Invariance} \label{sec:li}

\subsection{Spin $0$ fields} 
The action functional for a complex scalar field $\phi$ put forth in \cite{mp03} is,
\begin{align}
S 
&= \int d^{4}x \mathcal{L}_{MP_\phi} \nonumber \\
&= \int d^{4}x \left[|\partial \phi|^2 - m^2|\phi|^2 \right] 
+ \int d^{4}x \frac{i\kappa}{M_{Pl}}\phi^*\partial_{n}^3\phi 
\label{eq:lmpscalar}\\
&= S_{S} + S_{V_S} , \nonumber
\end{align} 
with $\kappa$ being a real, dimensionless parameter and $\mathbf{n}\cdot \mathbf{\partial} \equiv \partial_n$. $S_{S}$ and $S_{V_S}$ denote respectively the standard action for a complex scalar field and the new Lorentz violating part. Under an infinitesimal Lorentz transformation, $\delta_{\alpha \beta} S_{S} = 0$ while
\begin{align}
\delta_{\alpha \beta} S_{V_S}
= \int d^4x \phi^*n_{[\alpha}\partial_{\beta]}\partial_{n}^2\phi~. \label{eq:actionlorv} 
\end{align}
On the other hand, if the spacetime divergence of the N\"other current $\mathbf{\mathcal{J}}$
corresponding to Lorentz transformations is computed, we get
\begin{align}
\partial_\mu \mathcal{J}^\mu_{\alpha\beta} = -
\frac{\partial \mathcal{L}_{MP_\phi}}{\partial n^\lambda}(\delta
n^\lambda)_{\alpha\beta} = n_{[\alpha}\frac{\partial
\mathcal{L}_{MP_\phi}}{\partial n^{\beta]}}
= \phi^*n_{[\alpha}\partial_{\beta]}\partial_{n}^2\phi~. \label{eq:notherlorv} 
\end{align}
If Lorentz transformations are symmetries of the system then we must simultaneously have $\delta_{\alpha \beta} S_{V_S} = 0$ and $\partial_\mu\mathcal{J}^\mu_{\alpha\beta} = 0 = n_{[\alpha}\frac{\partial
\mathcal{L}_{MP}}{\partial n^{\beta]}} $. Requiring either of these yields the condition 
$n_{[\alpha}\partial_{\beta]}\partial_{n}^2\Phi = 0$.

A possible non-trivial solution is,
$\partial_{n}^2\phi = f(\mathbf{x}.\mathbf{n}) = f(z)$
where $z \equiv \mathbf{x.n}$. It is convenient in flat
spacetime to resolve the coordinate 4-vector along and orthogonal to $\mathbf{n}$:
\begin{align*}
\mathbf{x} = \mathbf{x}_{\parallel} + \mathbf{x}_{\perp}
\end{align*}
where $\mathbf{n}\cdot \mathbf{x}_{\perp} = 0$ and 
$\mathbf{x}_{\parallel} = \frac{\mathbf{x}\cdot \mathbf{n}}{n^2}~\mathbf{n} = \frac{z}{n^2}\mathbf{n}$. So, the derivative operator can be written as,
\begin{align*} 
\mathbf{\partial} 
&= \mathbf{\partial}_{\parallel} + \mathbf{\partial}_{\perp}, \\
&= \mathbf{n}\partial_z + \mathbf{\partial}_{\perp}
\end{align*}
where for ease of notation $\mathbf{\partial}_{\parallel} \equiv \mathbf{\partial}_{\mathbf{x}_{\parallel}}$,
$\mathbf{\partial}_{\perp} \equiv \mathbf{\partial}_{\mathbf{x}_{\perp}}$.
It is straightforward to show that
$\partial_n \phi(\mathbf x) = n^2 \partial_{z} \phi(\mathbf{x}_{\parallel},\mathbf{x}_{\perp})$
This implies in its turn
\begin{align}
\phi(\mathbf x) = \phi_{\parallel}(\mathbf{x}_{\parallel}) + \phi_{\perp}(\mathbf{x}_{\perp}) \label{eq:phifinal}
\end{align} 
where, $\phi_{\parallel}$ and $\phi_{\perp}$ are arbitrary functions of their
arguments. Under a Lorentz transformation $\mathbf{x}_{\parallel}$ and $\mathbf{x}_\perp$ will mix because thay have not been defined in a Lorentz invariant fashion. Hence, the imprint of Lorentz violation introduced into the action is borne by the decomposition of the scalar field to ensure that the action retains Lorentz symmetry.

If $\mathbf{n}$ is timelike, we can choose
coordinates such that $x^0$ lies along $\mathbf{n}$. Then our condition (\ref{eq:phifinal})
implies that when the full scalar field is a linear combination of a
time-dependent, spatially homogeneous piece and a static spatially
inhomogeneous piece, {\it the theory will possess Lorentz symmetry}. As a matter of fact, these fields indeed provide a representation of Lorentz algebra as we hope to show in our future publication.

\subsection{Maxwell (spin 1) field}
In this case, the usual kinetic term of the free
Maxwell field  and a dimension five, $\mathbf{n}$ dependent
operator constitute the modified Lagrangian density proposed in \cite{mp03},
\begin{align}
S = \int d^4 x \left[-\frac{1}{4}F_{\mu \nu}F^{\mu \nu} 
+ \frac{\xi}{M_{Pl}}n^{\mu}F_{\mu \nu}n^{\alpha}\partial_{\alpha}n_{\rho}\tilde{F}^{\rho \nu}\right]
\label{eq:lmpa}
\end{align}
where $\xi$ is a dimensionless parameter constraining Lorentz violation. For
convenience, we define $n^\mu F_{\mu \nu} \equiv F_{n \nu}, n_{\rho}
\tilde{F}^{\rho \nu} = \tilde{F}^{n \nu}$. Directly studying the variation of the action or the divergence of
the N\"{o}ther current of Lorentz transformation in parallel with the argument
given in case of the scalar field, the condition for the theory to be Lorentz
invariant is
\begin{align}
n_{[\alpha}F_{\beta]\nu}\partial_{n}\tilde{F}^{n\nu} 
+ F_{n\nu}n_{[\alpha}\partial_{\beta]}\tilde{F}^{n\nu}
+ F^{n\nu}\partial_{n}n_{[\alpha}\tilde{F}_{\beta]\nu}
= 0~. \label{eq:acon}
\end{align}
The last term is always zero because
$F^{n\nu}\partial_{n}n_{[\alpha}\tilde{F}_{\beta]\nu} =
F^{n\nu}\partial_{n}n_{[\alpha}\epsilon_{\beta]\nu \lambda \sigma}F^{\lambda
\sigma}$ contains a fifth rank completely antisymmetric tensor
($\tilde{F}^{\mu \nu} = \frac{1}{2}\epsilon^{\mu \nu \lambda \sigma}F_{\lambda \sigma}$)
in four spacetime dimensions. 




We transform to the Lorentz frame defined by $\mathbf{n} = (1,\vec{0})$, to
get a better physical picture of the problem in terms of electric and
magnetic field 3-vectors identified as $E_i = F_{0 i} =
\frac{1}{2}\epsilon_{lmi}\tilde{F}_{lm}, B_i = \tilde{F}_{0i} =
-\frac{1}{2}\epsilon_{ijk}F_{jk}$ where we have used $\epsilon^{0ijk} =
\epsilon_{ijk}$. The condition for Lorentz invariance becomes,
\begin{align}
\epsilon_{ijm}\partial_0 B_j B_m - E_j \partial_i B_j = 0 \label{eq:eb}  \\
\dot{\vec B} \times \vec B - \vec{\nabla}\vec B \cdot \vec{E} 
= 0 \label{eq:veceb} 
\end{align}
It is easy to see that if the fields are harmonic functions of spacetime as 
\begin{align}
\vec{E}=Re(\vec{\mathcal{E}}_0 exp(-iwt + i\vec{k}\cdot \vec{x}) ) \label{eq:efinal} \\
\vec{B}=Re(\vec{\mathcal{B}}_0 exp(-iwt + i\vec{k}\cdot \vec{x}) ) \label{eq:bfinal}
\end{align} 
they satisfy (\ref{eq:veceb}) when the relation $\vec{E}\cdot \vec{B} =
0$ (deduced from Bianchi identity) is incorporated. This ensures that it is
possible to have Lorentz symmetric electromagnetic fields in the modified
electrodynamics of \cite{mp03}.

\subsection{Spin $\frac{1}{2}$ field}
In \cite{mp03}, the action describing a Dirac spinor has been modified to,
\begin{align}
S &= \int d^4 x \
\overline{\psi}[(i\mathbf{\gamma}.\mathbf{\partial} - m)\psi
+ \frac{\mathbf{\gamma}.\mathbf{n}}{M_{Pl}}(\eta_{1} 
+ \eta_{2}\gamma_{5})\partial_{n}^2\psi]  \nonumber \\
&\equiv S_D + S_{V_D} \ , \label{eq:diracactn}
\end{align}
where $S_D$ is the standard Dirac action of a spinor field $\psi$ and
$S_{V_D}$ accounts for Lorentz violation. The dimensionless parameters
$\eta_1, \eta_2$ give the measure of Lorentz violation.

The only source of Lorentz violation is, by assumption, the appearance of the
constant $4$-vector $\mathbf{n}$ in $S_{V_D}$. Thus, there are no constant vectors
in the theory independent of $\mathbf{n}$. It is straightforward to show that,
under an infinitesimal Lorentz transformation, the action S changes by,
\begin{align}
\delta_{\alpha\beta}S_{V_D}
&= \frac{1}{M_{Pl}}
\int d^4 x \
\overline{\psi}
\{\mathbf{n}_{[\alpha}\gamma_{\beta]}
(\eta_{1} + \eta_{2}\gamma_{5})
\partial_{n}^{2}\psi \nonumber \\
&+ \mathbf{\gamma}.\mathbf{n}
(\eta_{1} + \eta_{2}\gamma_{5})
n_{[\alpha}\partial_{\beta]}\partial_{n}\psi   \label{eq:deltasv}
\end{align}
If we set 
$\partial_{n}\psi = \chi(z)$ , where $z=\mathbf{x}.\mathbf{n}$, then the second term in (\ref{eq:deltasv})
vanishes. After a partial integration (dropping the surface term), the first
term reduces to,
\begin{align}
\delta_{\alpha\beta}S_{V_D}
= - \frac{1}{M_{Pl}} 
\int d^4 x
\left[\eta_1 n_{[\alpha}\overline{\chi}\gamma_{\beta]}\chi
+ \eta_2 n_{[\alpha}\overline{\chi}\gamma_{\beta]}\gamma_{5}\chi \right] \nonumber \\
= - \frac{1}{M_{Pl}} 
\int d^4 x
\left[\eta_1 n_{[\alpha}J_{\beta]}(z)
+ \eta_2 n_{[\alpha}J_{\beta]}^{5}(z) \right] \ , \label{eq:currentdef}
\end{align}
where $J_{\alpha}(z) \equiv \overline{\chi}\gamma_{\alpha}\chi$,
$J_{\alpha}^{5}(z) \equiv \overline{\chi}\gamma_{\beta]}\gamma_{5}\chi$ etc.

Now, one can decompose the currents $\mathbf{J}$ and $\mathbf{J}^5$ as
$\mathbf{J} = \left(\frac{\mathbf{n}.\mathbf{J}}{n^2}\right)\mathbf{n} +
\mathbf{J}_{\perp} $ and $\mathbf{J}^5 =
\left(\frac{\mathbf{n}.\mathbf{J}^5}{n^2}\right)\mathbf{n} +
\mathbf{J}^5_{\perp} $ where $\mathbf{n}.\mathbf{J}_{\perp} = 0,\
\mathbf{n}.\mathbf{J}^5_{\perp} = 0$. Inserting this decomposition into
(\ref{eq:currentdef}), it is clear that
\begin{align}
\delta_{\alpha\beta}S_{V_D}
= -\int d^4 x
\left[\eta_1 n_{[\alpha}J_{\perp,\beta]}(z)
+ \eta_2 n_{[\alpha}J_{\perp,\beta]}^{5}(z) \right] \ , \label{eq:haha}
\end{align}
so that Lorentz violation now depends on the current $4$-vectors $\mathbf{J}_{\perp}$ and $\mathbf{J}^5_{\perp}$.

It should be noted, however, that \textit{these current 4-vectors are
orthogonal to $\mathbf{n}$ and are constants in the direction they point!} If,
for example, $\mathbf{n}$ is timelike, the currents $\mathbf{J}_{\perp}$ and
$\mathbf{J}^5_{\perp}$ must be spacelike and yet must be \textit{spatially
homogeneous}, being functions of $z$. This makes them constant 4-vectors
\textit{independent} of $\mathbf{n}$. Since, by assumption there are no
constant 4-vectors in the problem apart from $\mathbf{n}$, these currents must
vanish. 

As illustrated for the scalar field, requirement of Lorentz invariance of the action implies that,
\begin{align}
\psi(\mathbf{x}) = \psi_{\parallel}(\mathbf{x}_{\parallel}) + \psi_{\perp}(\mathbf{x}_{\perp}).\label{eq:psifinal}
\end{align}
In the preferred frame $\mathbf{n} = (1,\vec{0})$, $\psi_{\parallel}(\mathbf{x}_{\parallel})$ is a
spatially homogeneous spinor whereas the spinor $\psi_{\perp}(\mathbf{x}_{\perp})$ is time independent.

Hence, the invariance of the action under infinitesimal Lorentz transformations in our case does indeed lead to nontrivial restrictions on the functional form of fields in that the fields decouple into two parts one of which will be function only of certain projections of the coordinate vector along the fixed vector $\mathbf{n}$, the other part being a function of projections of the coordinate vector orthogonal to $\mathbf{n}$. These constraints have not been imposed by hand: they are the most
general solutions of the equations that result on requiring invariance
under infinitesimal Lorentz transformations. 


\section{Evaluation of Dispersion Relation}\label{sec:dr}
Now that we have found quite general and non-trivial field
configurations that make the modified scalar, vector and spinor actions of \cite{mp03} Lorentz invariant, the next step will entail calculating the
dispersion relations obeyed by these special fields.

\subsection{Scalar field}
The scalar field $\phi(x)$
assumed to be given by (\ref{eq:phifinal}) leads to the equation of motion
$(\Box + m^2) \phi = \frac{i\kappa}{M_{Pl}} \partial_n^3 \phi $ to be written as 
\begin{align}
(\mathbf{\nabla}_{\perp}^2 + m^2) \phi_{\perp} = - (\mathbf{\nabla}_{\parallel}^2 
+ m^2) \phi_{\parallel}  
+ \frac{i \kappa}{M_{Pl}}\partial_n^3 \phi_{\parallel}
  ~ \label{eq:seom}
\end{align}
Here, we have used the decomposition,
\begin{align*}
\Box \phi = \mathbf{\nabla}_{\parallel}^2 \phi_{\parallel} + 
\mathbf{\nabla}_{\perp}^2 \phi_{\perp}
\end{align*}
when $\mathbf{n}$ is a unit vector. It is obvious that to make sense of (\ref{eq:seom}) we must set both sides to a
constant which we choose to vanish for convenience. By taking the simple ans\"atz
$\phi_{\perp} \sim \exp (-\mathbf{k}_{\perp} \cdot \mathbf{x}_{\perp})$ and
$\phi_{\parallel} \sim \exp (-i~\mathbf{k}_{\parallel} \cdot \mathbf{x}_{\parallel})$, it is easy to see that the dispersion relations for the fields $\phi_{\parallel}$ and $\phi_{\perp}$ are respectively
\begin{align}
\mathcal E_{\parallel}^2 
&= |\vec{k_{\parallel}}|^2 + m^2 
+ \frac{\kappa}{M_{Pl}} (\mathbf{n}\cdot \mathbf{k}_{\parallel})^3 , \label{eq:phipardr} \\
\mathcal E_{\perp}^2 
&= |\vec{k_{\perp}}|^2 - m^2 . \label{eq:phiperpdr}
\end{align}
If we go to the inertial frame where $\mathbf{n}=(1,\vec{0})$ the dispersion relations take the simplified forms:
$\mathcal E^2 = m^2 + \frac{\kappa}{M_{Pl}} ~E^3~~ ; ~~ |\vec{k}|^2 = m^2$ provided the four momentum $\mathbf{k} = (\mathcal E, \vec{k})$. One can now eliminate $m^2$ from these equations to get the dispersion relation of the complete scalar field $\phi(\mathbf{x})$ in the high energy regime $\mathcal E \simeq |\vec{k}| >> m$,
\begin{align}
\mathcal E^2 \simeq |\vec{k}|^2 + \frac{\kappa}{M_{Pl}} ~|{\vec k}|^3
~\label{eq:liscaldisp}
\end{align}
which is same as the dispersion relation (\ref{eq:scalardisp}) computed in \cite{mp03}.

\subsection{Vector field}
The equation of motion obtained by the variation of the
action (\ref{eq:lmpa}) is (derived in \cite{monturrutia05}),
\begin{align}
\partial_{\mu}F^{\mu \nu} + \frac{\xi}{M_{Pl}}\left(n_{\rho}\epsilon^{\rho
    \sigma \mu \nu}\partial_{\mu}\partial_n F_{n \sigma} - \partial_n^2 \tilde{F}^{n\nu} \right) = 0 \label{eq:emeom}
\end{align}
The above equation and the Bianchi identity $\partial_{[\mu}F_{\nu \rho]}=0$
in the chosen reference frame are equivalent to the following equations:
\begin{align}
\vec{\nabla}\cdot \vec{E} 
= &0 
= \vec{\nabla}\cdot \vec{B} \label{eq:bianchivec} \\
\vec{\nabla}\times \vec{E} 
&= -\partial_t \vec{B} \label{eq:faraday} \\
-\dot{\vec{E}} + \vec{\nabla} \times \vec{B} 
&+ \frac{\xi}{M_{Pl}}(\ddot{\vec{B}} 
- \vec{\nabla} \times \dot{\vec{E}}) = 0 \label{eq:maxmod}
\end{align}
These are the modified free Maxwell equations. If we take the curl of both
sides of (\ref{eq:maxmod}), simplify using the Bianchi identity
(\ref{eq:bianchivec}), substitute the LI solution for the magnetic field and
assume $\mathbf{k} = (\omega,0,0,k^3)$, the dispersion relations at high
energy $\omega \simeq |\vec{k}|$ are:
\begin{align}
\omega_{R,L}^2 - |\vec{k}|^2 \simeq \pm \frac{2\xi}{M_{Pl}}|\vec{k_\perp}|^3 . \label{eq:livecdisp} 
\end{align}
The plus and minus signs appear for right and left circularly polarised
electromagnetic waves respectively.

\subsection{Spinor field}
If we take the spacetime dependance of the fields in (\ref{eq:psifinal}) to be
$\psi_{\parallel} \sim exp(-i~\mathbf{k}_{\parallel}\cdot \mathbf{x}_{\parallel})$, 
$\psi_{\perp} \sim exp (-\mathbf{k}_\perp \cdot \mathbf{x}_{\perp})$
and proceed as for the scalar field then the dispersion relations for the fields $\psi_{\parallel}$ and $\psi_{\perp}$ turn out respectively to be, 
\begin{align}
\left[\mathcal E_{\parallel}^2 
- |\vec{k_{\parallel}}|^2 - m^2 
- \frac{2 (\mathbf{n}\cdot \mathbf{k}_{\parallel})^3 }{M_{Pl}}(\eta_1 + \eta_2 \gamma_5)\right]
\psi_{\parallel}
= 0, 
\label{eq:psipardr}
 \\
\left[\mathcal E_{\perp}^2 
- |\vec{k_{\perp}}|^2 
+ m^2\right]\psi_{\perp}
= 0 . 
\label{eq:psiperpdr}
\end{align}
We are interested in high energy phenomena and at sufficiently high energies the massive spinors can be treated as chirality operator eigenstates. Redifining $\eta_{R,L} \equiv \eta_1 \pm \eta_2$, the dispersion relations in the special Lorentz frame we have selected in earlier instances simplify to 
$\mathcal E^2 - m^2 - \frac{2 ~\mathcal E^3}{M_{Pl}}\eta_{R,L} = 0$, 
$|\vec{k}|^2 - m^2 = 0$. Here, the four momentum of field $\psi(\mathbf x)$ is $\mathbf{k} = (\mathcal E, \vec{k})$. We can combine these two equations and get the dispersion relation of $\psi(\mathbf x)$ :
\begin{align}
\mathcal E^2 
- |\vec{k}|^2 - \frac{2 |\vec k|^3}{M_{Pl}}\eta_{R,L} 
\simeq 0
\label{eq:lispindisp}
\end{align}
in the limit of high energy $\mathcal E \simeq |\vec{k}| >> m$.

We have demonstrated invariance for tranformations only close to the identity in the
parameter space of the Lorentz group. However, in dealing with Lorentz tranformations in field theory, one is invariably dealing with the simply-connected universal cover $\mathop{\rm SL}(2,\mathbb{C})$ of the Lorentz group and so it is quite adequate to consider only the Lorentz Lie
Algebra and its action on the fields. Indeed, the Lorentz
Lie Algebra is realized on our special field configurations because the N\"other current appropriate to the transformations is conserved and also the full action is invariant under infinitesimal tranformations. This should guarantee invariance under the connected part of the Lorentz group which is of concern here provided we adhere to the special field configurations. In fact, this is sufficient to obtain the non-standard dispersion relations as well, as has been illustrated. 

The subject of our next project is to explicitly exhibit the closure property of Lorentz algebra as realised on the Lorentz preserving fields (manuscript under preparation). The first step towards this is to appropriately choose the generalised coordinates and construct the canonically conjugate momenta for the higher derivative Myers Pospelov model. It appears that one can set up a well posed initial value formulation of this theory with respect to our field solutions, thus eliminating the possibility of appearance of ghost states (auxilliary degrees of freedom that do not contribute towards the dynamics).


\section{Possible Application in Cosmology} \label{sec:application}
The field configurations that we have obtained have aspects of intrinsic interest when one considers prospective
application to cosmology as in inflationary scenarios. A peculiarity of almost all models of inflation \cite{niemeyer01, lemoine01} is that the phase of accelerated expansion lasted long enough for present scales of cosmological interest to be redshifted from trans-Planckian length scales at the onset of inflation. Hence, Lorentz violation must have been an intrinsic feature of the nascent Universe. Here again, one of the phenomenological approaches towards studying this era consists of modifying the standard dispersion relation of the scalar inflaton field together with the introduction of a standard timelike unit vector in the effective Lagrangian to define the preferred frame. However, the altered dispersion relation must obviously reduce to the standard linear behaviour at energies much smaller than $M_{Pl}$. The Myers and Pospelov theory shares these criteria which make it suitable for studying the physics of the Universe at the beginning of inflation. The fact that there is
a natural decomposition in Lorentz-preserving (scalar) fields between
spatially homogeneous and inhomogeneous parts implies that while the former,
in a Friedmann-Robertson-Walker (FRW) background spacetime, can play the role of the
inflaton field, the latter, acting as a perturbation on the former, may
provide natural seeds for the growth of inhomogeneities in the Universe. Moreover, in Friedmann-Robertson-Walker spacetime, a chosen frame exists by construction. Hence, the constant vector $\mathbf{n}$ can be taken to be orthogonal to the homogeneous isotropic spatial sections in FRW spacetime such that $\mathbf n = (1,\vec 0)$ in the comoving frame. 

Just as an illustration, we demonstrate how inhomogeneities appear in the energy momentum tensor in a Minkowski spacetime where $T_{\mu \nu}
= \frac{\partial \mathcal{L_{MP_\phi}}}{\partial(\partial_\mu \phi)} \partial_\nu \phi - \eta_{\mu\nu}\mathcal{L_{MP_\phi}}$. For fields of the form (\ref{eq:phifinal}) described by the Lagrangian density (\ref{eq:lmpscalar}), the energy density $\rho$ and pressure p  in the massless limit as measured in the chosen frame are:
\begin{align}
\rho &= T_{00} \nonumber \\
&= \dot{\phi_\parallel}^2 - \frac{i\kappa}{M_{Pl}}\phi_\parallel^*\dddot{\phi_\parallel}
 - \frac{i\kappa}{ M_{Pl}}\phi_\perp^* \dddot{\phi_\parallel} + (\vec{\nabla} \phi_\perp)^2\label{eq:energydensity} \\
p &= -\frac{1}{3} T_{ii} \nonumber \\
&= \dot{\phi_\parallel}^2 + \frac{i\kappa}{M_{Pl}}\phi_\parallel^*\dddot{\phi_\parallel} 
+ \frac{i\kappa}{ M_{Pl}}\phi_\perp^* \dddot{\phi_\parallel} 
- \frac{1}{3} (\vec{\nabla} \phi_\perp)^2
\label{eq:pressure}
\end{align}
The additional second and third terms in the energy density $\rho$ which follow from the new term in the action, may serve as sources of perturbations on the flat Minkowski metric when gravity is considered. This indicates that presence of Lorentz violation in an originally flat spacetime will automatically introduce non-trivial curvature.
The inhomogeneous field $\phi_\perp (\mathbf{x}_\perp)$ appears only in the last two terms of the expression for $\rho$ which can be understood as perturbations over the homogeneous
energy density of the field $\phi_\parallel (\mathbf{x}_\parallel)$. The Lorentz-preserving perturbations
due to the field $\phi_\perp ({\mathbf x_\perp})$ in \ref{eq:energydensity}, \ref{eq:pressure} might lead to growth of
Lorentz invariant inhomogeneities in spacetime. 

\cite{kanno06} provides an interesting exposition on the impact of Lorentz violation on the inflationary scenario in the scalar-vector-tensor model of gravity, where the Lorentz violating vector $u^\mu$ (as per the notation of \cite{kanno06}) is constrained to be unit and timelike. The authors here have deduced that the preferred frame, determined by $u^\mu$, practically aligns with the Cosmic Microwave Background rest frame and parameterised the metric for a homogeneous isotropic spacetime as
\begin{align*}
ds^2 = \mathcal{N}^2 (t) dt^2 - e^{2\alpha (t)} \delta_{ij}dx^{i}dx^{j}
\end{align*}
The lapse function $\mathcal{N} (t)$ also appears in the components of $u^\mu$: $u^\mu = (\frac{1}{\mathcal N}, \vec{0})$. Obviously, this choice of the metric and Lorentz violating vector is more general and far more appropriate for understanding cosmological aspects than Minkowski spacetime. Our cosmological example above is a special case where $\mathcal N = 1, \alpha = 0$. \cite{kanno06, bazeia06, bazeia08, avelino09} illustrate how Lorentz violating inflationary solutions for a family of models can be found even in the absence of any inflationary potential.   

True significance of our Lorentz preserving fields in the study of inflation, particulary Trans-Planckian modes, will be realised only when we suitably modify the Myers Pospelov effective action to hold in a spacetime with non-trivial curvature. We are interested in exploring how the inhomogeneous term appearing in the energy density may lead to the Jeans instability or some other aspect of structure formation in an FRW background.  We hope to report on this in future. 

\acknowledgements
We are indebted to F.P. Schuller for bringing to our notice \cite{schuller10} which gives a mathematical characterisation of a dispersion relation in any background geometry in terms of functions on the cotangent bundle based on the fundamental physical assumptions of predictivity and an observer independent notion of positive energy. We also thank the referees for suggesting improvements to the manuscript.


\begin{thebibliography}{0}

\bibitem{carroll90}
	\Name{Carroll S. M., Field G. B. \and Jackiw R.}
	\REVIEW{Phys. Rev. D}{41}{1990}{1231}.

\bibitem{latorre95}
	\Name{Latorre J. I., Pascaul P. \and Tarrach R.}
	\REVIEW{Nucl. Phys. B}{437}{1995}{60}.

\bibitem{colladay97}
	\Name{Colladay D. \and Kostelecky V. A.}
	\REVIEW{Phys. Rev. D}{55}{1997}{6760}.

\bibitem{amelino98}
	\Name{Amelino-Camelia G. \etal.}
	\REVIEW{Nature}{393}{1998}{763}.

\bibitem{colladay98}
	\Name{Colladay D. \and Kostelecky V. A.}
	\REVIEW{Phys. Rev. D}{58}{1998}{116002}.
  
\bibitem{coleman99}
	\Name{Coleman S. \and Glashow S.}
	\REVIEW{Phys. Rev. D}{59}{1999}{116008}.

\bibitem{gambini99}
	\Name{Gambini R. \and Pullin J.}
	\REVIEW{Phys. Rev. D}{59}{1999}{124021}.

\bibitem{schaefer99}
	\Name{Schaefer B. E.}
	\REVIEW{Phys. Rev Lett.}{82}{1999}{4964}.

\bibitem{biller99}
	\Name{Biller S. D. \etal.}
	\REVIEW{Phys. Rev. Lett.}{83}{1999}{2108}.

\bibitem{kifune99}
	\Name{Kifune T.}
	\REVIEW{Astrophys. J. Lett.}{518}{1999}{L21}.

\bibitem{amelino99}
	\Name{Amelino-Camelia G.}
	\REVIEW{Nature}{398}{1999}{216}.

\bibitem{ellis00}
	\Name{Ellis J. R. \etal.}
	\REVIEW{Astrophys. Jou.}{535}{2000}{139}.

\bibitem{urrutia00}
	\Name{Alfaro J., Morales-Tecotl H. A. \and Urrutia L. F.}
	\REVIEW{Phys. Rev. Lett.}{84}{2000}{2318}.

\bibitem{piran01}
	\Name{Amelino-Camelia G. \and Piran T.}
	\REVIEW{Phys. Rev. D}{64}{2001}{036005}.

\bibitem{kostelecky01} 
	\Name{Kostelecky, V. A. \and Mewes, M.} 
	\REVIEW{Phys. Rev. Lett.}{87}{2001}{251304}.

\bibitem{kostelecky02} 
	\Name{Kostelecky, V. A. \and Mewes, M.}
	\REVIEW{Phys. Rev. D}{66}{2002}{056005}.

\bibitem{sarkar02} 
	\Name{Sarkar S.} 
	\REVIEW{Mod. Phys. Lett. A}{17}{2002}{1025}.

\bibitem{alfaro02} 
	\Name{Alfaro J., Morales-Tecotl H. A. \and Urrutia L. F.}
	\REVIEW{Phys. Rev. D}{65}{2002}{103509}.

\bibitem{alfaro026} 
	\Name{Alfaro J., Morales-Tecotl H. A. \and Urrutia L. F.}
	\REVIEW{Phys. Rev. D}{66}{2002}{124006}.

\bibitem{ellis03} 
	\Name{Ellis J. R., Mavromatos N. E., Nanopoulos D. V. \and  Sakharov A. S.}
	\REVIEW{Astron. Astrophys.}{402}{2003}{409}.

\bibitem{jacob03} 
	\Name{Jacobson T., Liberati S. \and Mattingly D.}
	\REVIEW{Nature}{424}{2003}{1019}.

\bibitem{jacob05} 
	\Name{Jacobson T., Liberati S. \and Mattingly D.}
	\REVIEW{Springer Proc. Phys.}{98}{2005}{83}. 

\bibitem{albert07} 
	\Name{Albert J. \etal (MAGIC Collab.)}
	\REVIEW{Astrophys. Jou.}{667}{2007}{358}.

\bibitem{albert072}
	\Name{Albert J. \etal (MAGIC Collab.)}
	\REVIEW{Astrophys. Jou.}{669}{2007}{862}.

\bibitem{albert08}
	\Name{Albert J. \etal (MAGIC Collab.)}
	\REVIEW{Phys. Lett. B}{668}{2008}{253}.

\bibitem{piran07} 
	\Name{Jacob U. \and Piran T.}
	\REVIEW{Nature Phys.}{3}{2007}{87}.

\bibitem{galaverni07} 
	\Name{Galaverni M. \and Sigl G.}
	\REVIEW{Phys. Rev. Lett.}{100}{2008}{021102}.

\bibitem{gibbons07} 
	\Name{Gibbons G. W., Gomis J. \and Pope C. N.}
	\REVIEW{Phys. Rev. D}{76}{2007}{081701}.

\bibitem{liberati08} 
	\Name{Maccione L. \and Liberati S.}
	\REVIEW{J. Cosmol. Astropart. Phys.}{08}{2008}{027}.
 
\bibitem{maccione08} 
	\Name{Maccione L. \etal} 
	\REVIEW{Phys. Rev. D}{78}{2008}{103003}.

\bibitem{kostelecky08} 
	\Name{Kostelecky V. A. \and Mewes M.}
	\REVIEW{Astrophys. Jou.}{689}{2008}{L1}.

\bibitem{kostelecky09} 
	\Name{Kostelecky V. A. \and Mewes M.}
	\REVIEW{Phys. Rev. D}{80}{2009}{015020}.

\bibitem{ellis09} 
	\Name{Ellis J., Mavromatos N. E. \and Nanopoulos D. V.}
	\REVIEW{Phys. Lett. B}{674}{2009}{83}.
 

\bibitem{smolin09} 
	\Name{Amelino-Camelia G. \and Smolin L.}
	\REVIEW{Phys. Rev. D}{80}{2009}{084017}.
 
\bibitem{abdo09} 
	\Name{Abdo A. \etal}
	\REVIEW{Nature}{462}{2009}{331}.

\bibitem{amelino09} 
	\Name{Amelino-Camelia G., Laemmerzahl C., Mercati F. \and Tino G. M.}
	\REVIEW{Phys. Rev. Lett}{103}{2009}{171302}.

\bibitem{gubitosi10}
	\Name{Gubitosi G., Genovese G., Amelino-Camelia G. \and Melchiorri A.}
	\REVIEW{Phys. Rev. D}{82}{2010}{024013}.

\bibitem{mp03} 
	\Name{Myers R. C. \and Pospelov M.}
	\REVIEW{Phys. Rev. Lett.}{90}{2003}{211601}.

\bibitem{matt05} 
	\Name{Mattingly D.}
	\REVIEW{Living Rev. Rel.}{8}{2005}{5}.

\bibitem{jacob06} 
	\Name{Jacobson T., Liberati S. \and Mattingly D.}
	\REVIEW{Ann. Phys.}{321}{2006}{150}. 

\bibitem{liberati09} 
	\Name{Liberati S. \and Maccione L.}
	\REVIEW{Annu. Rev. Nucl. Part. Sci.}{59}{2009}{245}.

\bibitem{zwan71} 
	\Name{Zwanziger D.}
	\REVIEW{Phys. Rev. D}{3}{1971}{880}.

\bibitem{monturrutia05} 
	\Name{Montemayor R. \and Urrutia L. F.}
	\REVIEW{Phys. Rev. D}{72}{2005}{045018}.

\bibitem{niemeyer01} 
	\Name{Niemeyer J. C. \and Parentani R.}
	\REVIEW{Phys. Rev. D}{64}{2001}{101301(R)}.

\bibitem{lemoine01} 
	\Name{Lemoine M., Lubo M., Martin J. \and Uzan J-P}
	\REVIEW{Phys. Rev. D}{65}{2001}{023510}. 
 
\bibitem{kanno06}
	\Name{Kanno S. \and Soda J.}
	\REVIEW{Phys. Rev. D}{74}{2006}{063505}.

\bibitem{bazeia06}
	\Name{Bazeia D., Gomes C. B., Losano L. \and Menezes R.}
	\REVIEW{Phys. Lett. B}{633}{2006}{415}.

\bibitem{bazeia08}
	\Name{Bazeia D., Losano L., Rodrigues J. J. \and Rosenfeld R.}
	\REVIEW{Eur. Phys. J. C}{55}{2008}{113}.

\bibitem{avelino09}
	\Name{Avelino P. P. \etal}
	\REVIEW{Phys. Rev. D}{79}{2009}{123503}.

\bibitem{schuller10}
	\Name{R\"{a}tzel D., Rivera S. \and Schuller F. P.}
	\REVIEW{Phys. Rev. D}{83}{2011}{044047}.



\end{thebibliography}
\end{document}